# Photometric and Spectroscopic Analysis of Five Am Stars

Gireesh Chandra Joshi

Department of Physics, Government Degree College, Kotdwar-Bhabar, Uttarakhand, India

**Email address**
gchandra.2012@rediffmail.com

**Abstract**

The photometric analysis of sample Am stars is carried out to determine the stellar characteristics and to constrain the stellar dynamics. The spectroscopic analysis of the studied Am stars confirms their general characteristics of Am stars. The available data on elemental abundances for HD 113878 and HD 118660 have shown different characteristics during different epochs of observations. The basic stellar parameters (mass, luminosity, radius, life time, distance, proper-motion, etc.) are also determined to identify the stellar habitat zones for earth like exoplanet. Such information is important to identify suitable planets for human settlement in the near future. In this connection, the tidal radius and boundaries of the habitable zone of each star have been computed to support the search of an extra-terrestrial life around them. Asteroseismic mass scale test shows greater stellar masses comparable to the solar mass.

## 1. Introduction

Overpopulation is the major issue in our Earth and a requirement of more resources and space for human habitat. There are several telescopes are working out to search resources and space from outside the Earth. In this connection, the Chemical composition of stars is important to understand the physical parameters as well as stellar evolution process. The physical parameters and evolutionary stage are needed to estimate the habitat zone of studied stellar system for searching the earth like Exo-planets. Therefore, the chemical peculiarity of stars is carried out to decide merit of habitat zone compare to the solar system. Here, the author represents analytic results of the metallic-lined Am stars.

Such stars are a well-defined subgroup of the chemically peculiar (CP) stars on the upper main sequence (MS) [1]. They show strong and/or weak absorption lines of certain heavy and rare-earth elements in their optical spectrum in comparison to the normal stars of similar spectral type. Their peculiarity is interpreted as atmospheric under-abundance and/or over-abundance of different chemical elements and is explained by the diffusion process [2]. The atmospheres of these stars have an inhomogeneous distribution of chemical elements. Such distribution is frequently found in the abundance spots of the stellar surfaces and clouds of chemical elements as concentrated at the certain depths along the stellar radius [3]. Am stars exhibit overabundance of the iron peak and heavy elements such as *Zn, Sr, Zr, Ba* and underabundance of the elements such as *Ca, Sc* [4].

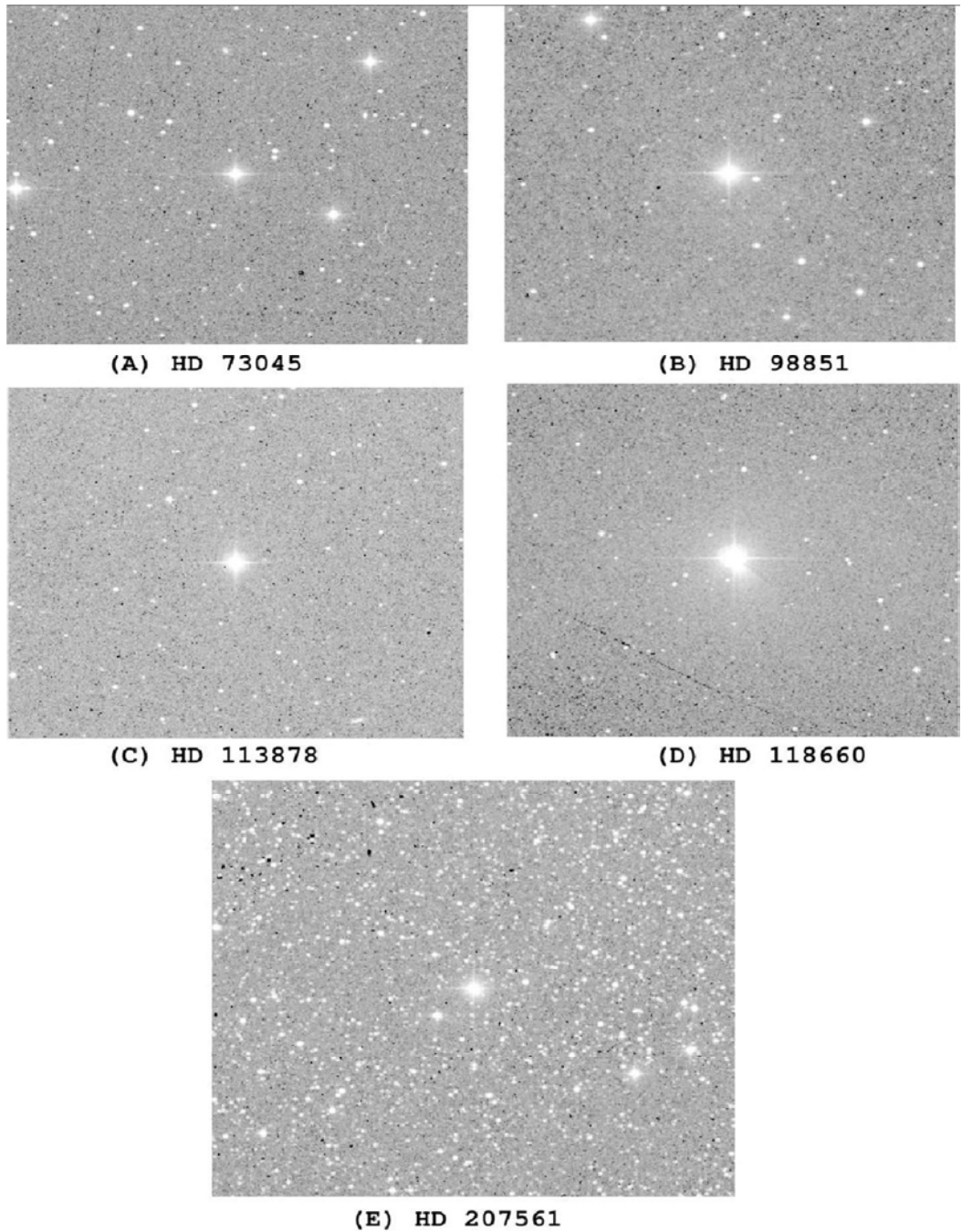

*Figure 1. The finding charts of HD 73045, HD 98851, HD 113878, HD 118660 and HD 207561 are shown in the panel A, B, C, D and E respectively. The size of each FC is $15 \times 15\ arcmin^2$ and constructed through "fits" file as extracted from the DSS1 server of ESO. These finding charts are effectively used to identify the target stars during their time series observation through fast photometer and CCD camera on the 1.04 m telescope of ARIES. The most centered and brighter star of each panel represents the corresponding target object.*

The strengths of metal lines of Am stars are more typical of the F star rather than an A star [5]. The Am stars are very weak magnetic and found generally in the binary system. They are also excellent test objects for astrophysical processes like diffusion, convection, and stratification in stellar atmospheres in the presence of strong magnetic fields [6]. The parametric values of these processes are used to constrain the characteristic properties of ejected radiation from studied star and constrain the impact of ejected radiation on the revolving planets within the habitat zone of studied stars.

## 2. Sample Stars and Their Previous Study

### 2.1. Characteristic Properties of Stars

In the present work, the author has taken a sample of Am field stars, having five members as HD 73045, HD 98851, HD 113878, HD 118660 and HD 207561. The present analysis is carried out by using existing catalogues and literature data. The detail of basic parameters of selected stars is given in the Table 1 and their corresponding finding charts (FCs) have been shown in the Figure 1. These FCs of target stars are helpful to decide the nature of observations as well as to perform the accurate analysis of gathered data. A brief discussion of available characteristic data of each sample star is given as below.

*Table 1. The detail of basic parameters of selected sample Am stars as extracted from the SIMBAD service. The distance of each star has been extracted from the McDonald et al. (2012).*

| Parameter Name | HD 73045 | HD 98851 | HD 113878 | HD 118660 | HD 207561 |
|---|---|---|---|---|---|
| $R.A.(\alpha)$ | $08^h36^m48^s$ | $11^h22^m51.17^s$ | $13^h06^m0.7^s$ | $13^h38^m08^s$ | $21^h48^m16^s$ |
| $Declination\ (\delta)$ | $+18°52'58"$ | $+31°49'41"$ | $+48°01'41"$ | $+14°18'07"$ | $+54°23'15"$ |
| $Distance\ (pc)$ | $218.34 \pm 0.21$ | $160.51 \pm 0.09$ | $401.61 \pm 0.31$ | $73.26 \pm 0.03$ | $116.69 \pm 0.07$ |
| $Radial\ Velocity\ (km/s)$ | $35.20 \pm 0.07$ | $-1.86 \pm 0.89$ | $-18.10 \pm 0.90$ | $-1.7 \pm 0.2$ | $-15.10 \pm 3.6$ |
| $Redshift\ z$ (order of $10^{-6}$) | 117 | $-0.6 \pm 0.3$ | $-6 \pm 0.3$ | $-0.06 \pm 0.03$ | $-50 \pm 12$ |
| $Parallaxes$ | $3.54 \pm 0.14$ | $6.64 \pm 0.06$ | $3.50 \pm 0.04$ | $14.18 \pm 0.05$ | $7.63 \pm 0.04$ |
| $Spectral\ Type$ | Am C | F2 D | A5 D | A9VsC | F0III C |

### 2.1.1. HD 73045

The spectral type of HD 73045 is given as A7.2 [7], Amvar [8] and Am [9]. This star belongs to the Praesepe open star cluster [10] and classified as Am star in which Ca and Sc are underabundant and the Fe-peak elements are overabundant [11]. The analysis of high spectral resolution and high signal-to-noise ratios observation gives the values of $T_{eff}$, $\log g$ and $Fe_H$ as 7500K, 4.00 and 0.35 respectively. Similarly, the high resolution spectroscopic work provides the values of $T_{eff}$, $\log g$ and $Fe_H$ as 7520K, 4.27 and 0.59 respectively [12]. Masana et al. 2006 [13] provides the value of $T_{eff}$ = 7524 K. Its velocity *(v)* given as 35.20±0.07 [14], 22.8 [15], 48.5 [16], 22.8±2.0 [17] and 22.8 [18].

### 2.1.2. HD 98851

It describes as F2 spectral type [19] and marginal Am stars [20]. Its velocity is found to be −6.60±3.50 [21]. Its luminosity class is designed to be F3III [22]. The proper motion values ($\mu_x$, $\mu_y$) of HD 98851 are $(-46 \pm 10, -21 \pm 10)$ [23]. The main pulsating frequencies of HD 98851 found to be 0.208 and 0.103 $mHz$ and its equivalent H-line spectral class is F1-IV [24].

### 2.1.3. HD 113878

HD 113878 is a A5 spectral type star with metallic lines [19] and Am star [22]. It shows $\delta$ − Sct behavior with a pulsation period of 2.31hr (f1 = 0.12mHz) [25]. The values of log g are given to be 3.36 [26] and 3.84 [27]. The value of [F e/H] is 0.74 for this star [26]. Its parallax values are 2.67±0.99 [28] and 2.49±0.77 [29]. The proper motion values ($\mu_x$, $\mu_y$) of HD 113878 are (−14±8, −25±8) [23]. Its velocity is $-18.10 \pm 0.9$ [21].

### 2.1.4. HD 118660

The spectral class of HD 118660 is A9Vs [30] and A9III-IV [31]. The calculated values of its velocity are −1.70±2.90 [21], -1.7 [15], -1.7 [17] and −1.7±2.0 [32]. It also shows $\delta$ − Sct behavior with multi-periods (1 hr, 2.52hr etc.) [25].

### 2.1.5. HD 207561

It has described to be F0III [33] or F0IV [34] spectral star. The period of HD 207561 is found to be 6min. The null results of Bz indicate that 6 min periodicity of HD 207561 through fast photometer and CCD photometry may be arising due to the instrumental or other reasons [6]. The possible rapid pulsation in the atmosphere of HD 207561 could have non-stable modes [35]. The rotational velocity (74±5kms$^{-1}$) of HD 207561 have been found through the analysis of high-resolution spectroscopic observations. Its proper motion values ($\mu_x$, $\mu_y$) are derived to be (+39±7, +40±8) [23]. Its velocity is −15.10±3.60 [21].

## 2.2. Previous Observations of High Resolution of Spectroscopy

The high-resolution spectroscopic data of HD 73045 have been acquired by Fossati et al. 2007 [11] through Echelle Spectro-Polarimetric Device for Observations of stars (ESPaDOnS) spectrometer at the Canada-France-Hawaii Telescope (CFHT) [11]. The ESPaDOnS spectrometer covers the wavelength range from 3700 Å to 10400 Å with a resolving power of about 65000. The high-resolution spectral data of HD 98851, HD 113878, HD 118660 and HD 207561 have acquired by Joshi et al. 2015 [6] and Joshi et al. 2012 [36]. The prescribed data were collected through the Nasmyth echelle spectrometer (NES) of the 6-m Big Telescope Alt-Azimuthal (BTA) telescope at the Special Astrophysical Observatory (SAO), Russian Academy of Sciences (RAS). It also includes the Main Stellar Spectrograph (MSS). The MSS has been applied to register the circularly polarized spectra within 500 Å range, have centered on 4700 Å. These instruments share the common block of preslit devices, including a set of calibration lamp. The mean resolving power $\lambda/\Delta_\lambda$ of about 14000 in spectropolarimetric mode and 39000 in Echelle spectroscopy. The detail of used spectra format and their quality is summarized in the Table 2.

*Table 2. High resolution spectroscopic observation log for the sample Am stars as extracted from the literature.*

| HD number | JD, 2450000+ | Instrument | Sp. Range (Å) | SNR | References |
|---|---|---|---|---|---|
| 73045 | 3745.091 | ESPaDOns | 6850-6800 | 290 (5000 Å) | [11] |

| HD number | JD, 2450000+ | Instrument | Sp. Range (Å) | SNR | References |
|---|---|---|---|---|---|
| 98851 | 5025.265 | NES | 4840-6250 | 250 (5500 Å) | [6] |
|  | 6383.463 | MSS | 4428-4980 | 370 (4700 Å) | [6] |
| 11383 | 5025.312 | NESS | 4840-6250 | 220 (5500 Å) | [6] |
|  | 5615.676 | SARG | 3600-7900 | 100 (5750 Å) | [39] |
|  | 6381.497 | MSS | 4428-4980 | 180 (4700 Å) | [6] |
|  | 6383.486 | MSS | 4428-4980 | 250 (4700 Å) | [6] |
| HD 118660 | 5615.649 | SARG | 3600-7900 | 100 (5750 Å) | [39] |
|  | 6408.505 | NESS | 3950-6860 | 130 (5500 Å) | [6] |
|  | 6383.521 | MSS | 4428-4980 | 350 (4700 Å) | [6] |
|  | 6383.538 | MSS | 4428-4980 | 330 (4700 Å) | [6] |
| 207561 | 5702.506 | NES | 4426-4978 | 230 (5500 Å) | [6] |
|  | 5764.342 | NES | 4420-4974 | 250 (5500 Å) | [6] |
|  | 5764.508 | NES | 4420-4974 | 200 (5500 Å) | [6] |

The abundances result of high-resolution spectroscopic observations of HD 113878 and HD 118660 are also available with the SARG (Spettrografo ad Alta Risoluzione del Galileo) spectroscope at the Telescopio Nazionale Galileo (TNG), La Palm [37]. SARG covers a spectral wavelength range from 3600 Å to 10000 Å, with a resolution ranging from 29000 to 164000 [38].

### 2.3. Hβ Profile of HD 113878 and Results for Other Stars

The absorption lines of Balmer series of the gathered stellar spectrum are effectively useful in the stellar astronomy due to their appearances in numerous stellar objects as a function of abundances of hydrogen. These lines are strong compared to other lines from other elements and used to determine radial velocity and surface temperature. The Hβ absorption line of Balmer series has a strong indicator of temperature. As a result, the stellar parameters such as $T_{eff}$ and $\log g$ have estimated through the best fit computed spectra with the observed spectrum of each sample star. In the Figure 2, the author represents the comparison of synthetic profile of Hβ line computed for $T_{eff}$ = 7000 K and log g = 3.7 with an observed spectrum of HD 113878 to show the best fitting of adopted synthetic Hβ profile on actual spectrum.

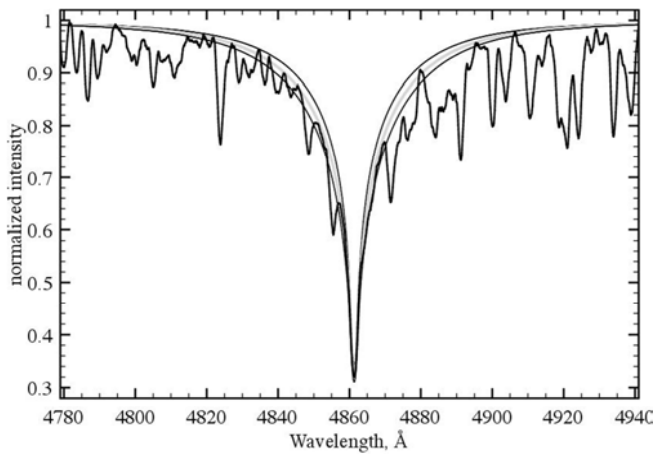

*Figure 2.* An observed Hβ line profile of HD 113878 and its result of fitting with synthetic spectrum as computed for the effective temperature of $T_{eff} = 7000 \pm 200\ K$ [6, 39].

Black thin lines of this figure represent the synthetic profiles as computed for the two values of $T_{eff}$ = 7000 ± 200K which are reflecting the accuracy of the procedure of determination of $T_{eff}$. These results are close to the results of Catanzaro & Ripepi 2014 [37], which are $T_{eff}$ = 6900±200K and log g = 3.4±0.1 dex. Similarly, Casagrande et al. 2011 [26] have derived the following astrophysical parameters for HD 113878 thorough the spectroscopic observations: $T_{eff}$= 7072±210K, log g = 3.36, and [Fe/H] = 0.74. The Hα and Hβ line wings of the high-resolution spectrum of HD 73045 are given the value of effective temperature ($T_{eff}$) and log g as 7570 K and 4.05 respectively [10]. Based on the low-resolution spectroscopy, the $T_{eff}$ and log g for the star HD 98851 are determined to be 7000±250K and 3.5±0.5 respectively [24]. The best fitted Hβ and Hδ profiles of the high-resolution spectrum of HD 118660 provide atmospheric parameters as $T_{eff}$ = 7200±200K, log g = 3.9±0.1 dex and $v \sin i$ = 100±10km s−1 [37]. After best fitted Hβ profile on the high-resolution spectrum of HD 207561, the values of $T_{eff}$ and surface gravity (log g) have found to be 7300±250 K and 3.7±0.1 dex respectively [36]. Based on the spectroscopic analysis of new gathered data, the values of a set ($T_{eff}$, log g) of HD 98851, HD 113878 and HD 118660 are found to be (7000±200, 3.65±0.15), (7000±200, 3.70±0.10) and (7550±150, 4.00±0.10) respectively [39].

## 3. Extraction of Stellar Magnitudes

The stellar magnitudes of selected Am field stars with an effective wavelength of filters have been listed in the Table 3. Author does not find the stellar magnitudes of HD 207561 in SDSS photometry. Similarly, HD 73045 and HD 113878 have also stellar magnitudes in filters of GALEX. Author has been noticed the descending stellar magnitudes with increment effective wavelength from filter z to filter R and stellar magnitude of G filter of each sample stars is greater than its stellar magnitude of R filter. In this connection, the stellar magnitudes further decrease with effective wavelengths from filter G to filter K. Author does not obtain any generalized feature of sample members below the effective wavelength of filter z and above the effective wavelength of filter K. Thus, author obtains a bump of stellar magnitude at the effective wavelength of filter G for each Am star of the sample.

Table 3. *Stellar Magnitudes in different filters under study of sample stars.*

| S. N. | Filter | λ | HD 73045 | HD 98851 | HD 113878 | HD 118660 | HD 207561 |
|---|---|---|---|---|---|---|---|
| 01 | FUV | 153.8 | 16.090 | ---- | 16.514 | ----- | ----- |
| 02 | NUV | 231.6 | 12.418 | ---- | 12.467 | ----- | ----- |
| 03 | u | 354.3 | 14.474 | 9.364 | 9.959 | 8.375 | ----- |
| 04 | g | 477.0 | 9.031 | 11.419 | 8.401 | 7.275 | ----- |
| 05 | r | 623.1 | 8.620 | 10.933 | 8.257 | 10.485 | ----- |
| 06 | i | 762.5 | 12.032 | 11.065 | 12.683 | 7.058 | ----- |
| 07 | z | 913.4 | 9.193 | 7.909 | 9.096 | 7.139 | ----- |
| 08 | B | 440.0 | 8.906 | 7.729 | 8.596 | 6.739 | 8.078 |
| 09 | V | 550.0 | 8.622 | 7.415 | 8.249 | 6.496 | 7.846 |
| 10 | R | 650.0 | 8.430 | 7.220 | 8.020 | 6.350 | 7.710 |
| 11 | G | 715.0 | 8.512 | 7.295 | 8.116 | 6.398 | 7.752 |
| 12 | J | 1235.0 | 8.036 | 6.741 | 7.581 | 5.973 | 7.377 |
| 13 | H | 1662.0 | 7.947 | 6.673 | 7.465 | 5.890 | 7.363 |
| 14 | K | 2159.0 | 7.944 | 6.610 | 7.407 | 5.853 | 7.314 |
| 15 | W1 | 3400.0 | 7.861 | 6.626 | 7.313 | 5.826 | 7.220 |
| 16 | W2 | 4600.0 | 7.887 | 6.549 | 7.402 | 5.750 | 7.327 |
| 17 | W3 | 12000.0 | 7.907 | 6.628 | 7.407 | 5.886 | 7.303 |
| 18 | W4 | 22000.0 | 7.587 | 6.530 | 7.156 | 5.815 | 7.122 |
| 19 | β | ----- | 2.759 | ----- | 2.745 | 2.778 | 2.825 |
| 20 | (b-y) | ----- | 0.218 | 0.199 | 0.219 | 0.150 | 0.142 |
| 21 | m1 | ----- | 0.195 | 0.222 | 0.257 | 0.214 | 0.220 |
| 22 | c1 | ----- | 0.749 | 0.766 | 0.728 | 0.794 | 0.819 |

Table 4. *The derived atmospheric parameters of the sample stars through the stromgreen photometry.*

| Parameter Name | HD 73045 | HD 98851 | HD 113878 | HD 118660 | HD 207561 |
|---|---|---|---|---|---|
| $B.C.$ | −0.7010 | −0.6618 | −0.7368 | −0.6247 | −0.5565 |
| $\log T_{Balona}$ | 4.088 | 4.079 | 4.094 | 4.075 | 4.062 |
| $\log g_{Balona}$ | 3.575 | 3.715 | 3.506 | 3.637 | 3.884 |
| $T_{MD}(K)$ | 7400 | 7500 | 7100 | 7500 | 7900 |
| $\log g_{MD}$ | 4.00 | 4.00 | 3.25 | 4.05 | 4.10 |
| $a_0$ | 0.2567 | 0.2436 | 0.1956 | 0.1792 | 0.1749 |
| $r^*$ | 0.0529 | 0.0478 | 0.0595 | 0.0544 | 0.0167 |
| $T_{eff}(K)$ | 7348 ± 59 | 7447 ± 60 | 7051 ± 55 | 7447 ± 60 | 7845 ± 63 |
| $\log g\ (cgs)$ | 3.85 | 3.86 | 3.13 | 3.91 | 3.97 |
| $M_V$ | 2.141 | 2.149 | 1.662 | 1.650 | 2.265 |

## 3.1. Atmospheric Parameters Through Stromgren Photometry

The values of temperature, logarithm gravity and BC are used to estimate the basic atmospheric parameters of any star. The values of bolometric corrections (BC), logarithm temperature and logarithm gravity are calculated through the Balona's relations [40] as follow,

$$BC = 0.2900 + 2.8467[c] + 2.8334[\beta] + 0.6481[c]^2 − 0.2997[\beta][c] + 2.1487[\beta]^2, \qquad (1)$$

$$\log T_{Balona} = 3.9036 − 0.4816[c] − 0.5290[\beta] − 0.1260[c]^2 + 0.0924[\beta][c] − 0.4013[\beta]^2, \qquad (2)$$

$$\log g_{Balona} = 5.9046 − 3.2262[c] + 4.0883[\beta] − 0.5383[c]^2 − 0.2774[\beta][c] − 0.0007[\beta]^2. \qquad (3)$$

The values of [β] and [c] are estimated as [β] = log (β − 2.5) and [c] = log (c + 0.2). The dereddened value ($a_0$) and reddening free parameters (r) are introduced to calibrate these parameters [41]. Theses introduced terms are defined as below,

$$a_0 = 1.36(b − y) + 0.36m + 0.18c − 0.2448 \text{ and } r^* = 0.35c − 0.07(b − y) − \beta + 2.565. \qquad (4)$$

These estimated parameters are used to estimate the value of $M_V = 1.5 + 6a + 17r$ [42], here $r^* = −r$. A synthetic ubvyβ indices to determine the values of effective temperature TMD and g in the range 600K < TMD ≤ 20000K through the grid procedure [41]. The calibrations of effective temperature ($T_{eff}$) and logarithm gravity are further carried out through the relation [43],

$$T_{MD}/T_{eff} = 1.007 \pm 0.008 \text{ and } \log g = \log g_{MD} - 2.9406 + 0.7224 \log T_{eff} \tag{5}$$

In the case of HD 98851, indices β misses in literature. In this connection, the value of r* of HD 98851 is estimated through a relation a = 2([m] − 0.179) + 0.8r* [44] and value of [m] is given as [m] =m + 0.30(b − y) [45]. Author found the values of [m], r* and synthetic β indices for star HD 98851 as 0.282, 0.0478 and 2.784 respectively. The estimated value of β indices of HD 98851 has utilized to estimate the $T_{eff}$ through above prescribed relations. All computed parameters of studied Am stars through the Stromgren photometry are listed in the Table 4. The values of $T_{eff}$ and *log g* of this table are the initial derived set of sample stars.

### 3.2. Measurement of Stellar Reddening

The absolute value of MV is also estimated through the relation $M_V = V + 5 + 5 \log \pi - A_V$, in which $\pi$ and $A_V$ are the stellar parallax and visual absorption. The values of visual absorption of nearby and high galactic latitude stars are estimated as $A_V = 3.1 \, E(B - V)$ [46] and $E(B - V)$ is the reddening value for studied star. The Hipparcos catalogue provides the parallax values of HD 73045, HD 98851, HD 113878, HD 118660 and HD 207561 as 4.97±1.22, 5.84±0.87, 2.67±0.99, 13.95±0.85 and 7.86±0.70 mas respectively (van Leeuwen, F., 2007). These values are provided the reddening values as -0.012, -0.37, -0.413, 0.19 and 0.75 for HD 73045, HD 98851, HD 113878, HD 118660 and HD 207561 respectively.

## 4. Abundance Analysis

### 4.1. Results of Spectroscopic Studies from Literature

The abundance of various elements is estimated through the spectrum synthesis method of the observed spectrum while the theoretical spectra were computed in LTE approach with Synth3 Software [47]. Analysis of stellar spectra is based on the principle that the different spectral lines of a certain chemical yield the same element abundances, independent of their equivalent width [48]. The spectroscopic abundance results of various elements of each star are depicted in the Figure 3 (A) and listed in the Table 5. The values of elemental abundances of SUN I and SUN II of Table 5 are extracted from the Joshi et al. 2015 [6] and Asplund et al. 2009 [49] respectively. As per the private communication of corresponding author with Honorable Eugene Semenko[1], different values of solar abundances of both studies arise due to the different normalization. Asplund et al. 2009 [49] used the abundances normalized to log(H)+12 to avoid negative values, while Eugene Semenko (in the work of Joshi et al. 2015) used a normalization log N(el)/N(tot) when the total sum of all abundances must equal to one. In the present work, average values of solar abundances of both studies have been utilized to construct the Figure 3. The high-resolution spectroscopic analysis of sample stars is showing the following features,

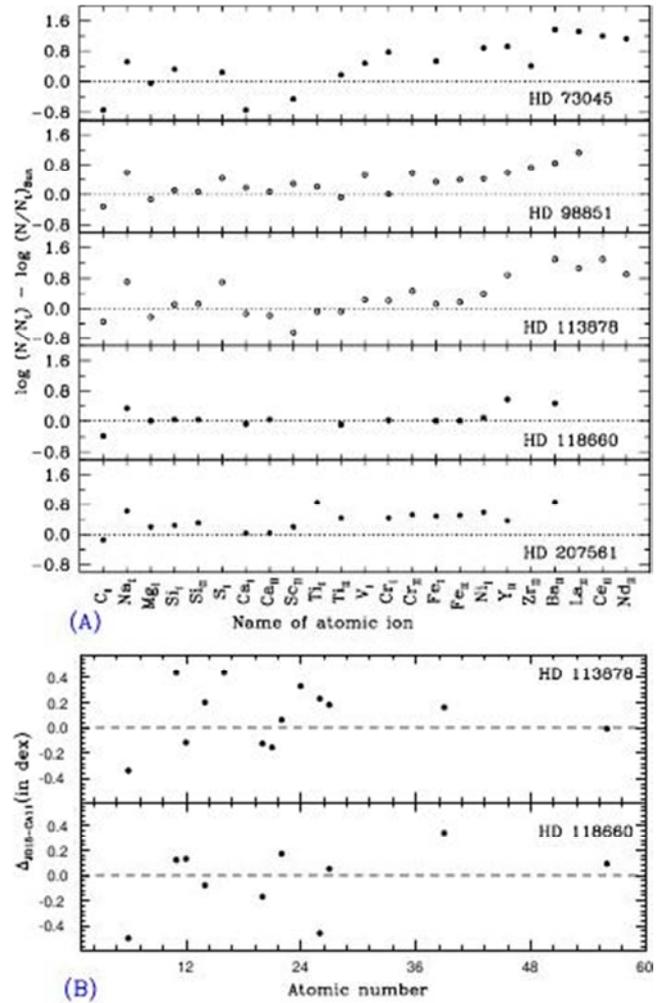

*Figure 3.* In the panel A): Derived atmosphere abundances of HD 98851, HD 113878 and HD 118660 by JO15 for the various elements relative to the Sun. Atmospheric abundance values for HD 73045 and HD 207561 are extracted from the Fossati et al. (2008) and Joshi et al. (2012) respectively. In the panel B): Author represents the difference of atmospheric abundances as derived by the Joshi et al. (2015) and Casagrande et al. (2011) with respect to the derived atmospheric abundances of the former work.

(1) The abundances of C, N, O, Ca and Sc are under-abundant for HD 73045. Other hand, Fe-peak and rare earth elements are found to be overabundant for it [10].

(2) Abundances of half of the elements HD 98851 are like solar one within error.

(3) In the case of HD 113878, author finds a deficiency in abundance of Ca, Sc, and Ti compare to solar values and excess in abundance of heavy elements.

(4) In the case of HD 118660, abundance of Ca and abundance of the group of lanthanide elements is found to be underabundant and overabundant respectively.

(5) The small underabundance of C, O, Ca, and Sc are reported for HD 207561 [6]. Similarly, the mild Am character and normal star abundances of HD 207561 have confirmed through moderate excess of the other 10 chemical elements (Na, Mg, Si, Ti, Cr, Mn, Fe, Ni, Y, Ba).

*Table 5. Individual abundance of chemical elements in the atmosphere of stars with solar chemical composition. The colon sign denotes doubtful measurements due to the small number of lines or others. The actual accuracy for most of the elements should be set up to 0.20-0.25 dex as depend in change in $\log N/N_{tot}$ caused by the accuracy of atmosphere parameters.*

| Ion | At. No. | HD 73045 FO08 | HD 98851 JO15 | HD 207561 JO15 | Sun I JO15 | Sun II AS09 | Sun |
|---|---|---|---|---|---|---|---|
| C I | 06 | −4.37 ± 0.04 | −3.94 ± 0.17 | −3.78 ± 0.10 | −3.61 | −3.65 | −3.63 |
| N I | 11 | −5.32 ± 0.03 | −5.25 ± 0.32 | −5.22 ± 0.19 | −5.80 | −5.87 | −5.84 |
| Mg I | 12 | −4.52 ± 0.04 | −4.59 ± 0.06 | −4.27 ± 0.03 | −4.44 | −4.51 | −4.48 |
| Mg II | 12 | | | −4.18 ± 0.00 | −4.51 | −4.51 | −4.51 |
| Si I | 14 | −4.21 ± 0.05 | −4.42 ± 0.12 | −4.29 ± 0.16 | −4.53 | −4.53 | −4.53 |
| Si II | 14 | | −4.45 ± 0.00 | −4.23 ± 0.00 | −4.53 | −4.53 | −4.53 |
| S I | 16 | −4.67 ± 0.09 | −4.47 ± 0.00 | | −4.91 | −4.90 | −4.91 |
| Ca I | 20 | −6.46 ± 0.04 | −5.53 ± 0.16 | −5.69 ± 0.24 | −5.70 | −5.73 | −5.72 |
| Ca II | 20 | | −5.64 ± 0.03 | −5.68 ± 0.00 | −5.70 | −5.73 | −5.72 |
| Sc II | 21 | −9.39 ± 0.01 | −8.65 ± 0.03 | −8.74 ± 0.10 | −8.89 | −8.99 | −8.94 |
| Ti I | 22 | | −6.91 ± 0.00 | −6.26 ± 0.37 | −7.09 | −7.14 | −7.12 |
| Ti II | 22 | −6.95 ± 0.04 | −7.19 ± 0.17 | −6.69 ± 0.29 | −7.09 | −7.14 | −7.12 |
| V I | 23 | −7.60 ± 0.00 | −7.55 ± 0.00 | | −8.11 | −8.04 | −8.08 |
| Cr I | 24 | −5.63 ± 0.16 | −6.38 ± 0.25 | −5.97 ± 0.07 | −6.40 | −6.40 | −6.40 |
| Cr II | 24 | | −5.82 ± 0.26 | −5.88 ± 0.09 | −6.40 | −6.40 | −6.40 |
| Fe I | 26 | −4.03 ± 0.04 | −4.23 ± 0.22 | −4.09 ± 0.15 | −4.54 | −4.59 | −4.57 |
| Fe II | 26 | | −4.17 ± 0.25 | −4.07 ± 0.10 | −4.54 | −4.59 | −4.57 |
| Ni I | 27 | −4.93 ± 0.06 | −5.39 ± 0.25 | −5.23 ± 0.18 | −5.82 | −5.81 | −5.82 |
| Y II | 39 | −8.91 ± 0.09 | −9.24 ± 0.15 | −9.47 ± 0.09 | −9.83 | −9.83 | −9.83 |
| Zr II | 40 | −9.05 ± 0.01 | −8.73 ± 0.00 | | −9.46 | −9.45 | −9.46 |
| Ba II | 56 | −8.50 ± 0.00 | −9.02 ± 0.22 | −9.00 ± 0.00 | −9.86 | −9.87 | −9.87 |
| La II | 57 | −9.61 ± 0.09 | −9.80 ± 0.20 | | −10.94 | −10.91 | −10.93 |
| Ce II | 58 | −9.26 ± 0.04 | | | −10.46 | −10.46 | −10.46 |
| Nd II | 60 | −9.48 ± 0.03 | | | −10.62 | −10.59 | −10.61 |

## 4.2. Dissimilarities of Analytic Results of Elemental Abundances

An individual discussion of elemental abundance can help to understand the atmospheric dynamics of Am stars. A typical analysis of the abundance of each element for HD 113878 and HD 118660 is carried out by both Joshi et al. (2015) [hereafter, JO15] and Casagrande et al. 2011 [hereafter, CA11]. The Author has found dissimilar properties of elemental abundances of some ion for HD 113878 and HD 118660 as shown in the Figure 3 (B). A brief description of dissimilar properties of ion abundances is given as below,

Carbon Abundances: - According to the data of Table 6 and Table 7, the value of carbon abundances of each star is found to be maximized. Based on the comparison of work JO15 and CA11, the difference of carbon abundance of HD 118660 is found to be maximum with respect to the other elements. However, these works also indicate that said difference for HD 113878 is larger except Na and S. JO15 reported that the carbon abundance is under-abundant for HD 113878, however it comes nearly equal to the solar one in the work of CA11. In the case of HD 118660, carbon abundance is found to be underabundance by JO15, whereas it reported to overabundant by CA11.

Magnesium Abundances: - JO15 found the Mg overabundant for HD 118660, whereas it is found to be under-abundance in the analysis of CA11.

Silicon Abundances: - CA11 found the Si overabundant for HD 113878, whereas it is found to be under-abundance in the analysis of JO15.

Calcium Abundances: - According to the CA11, the elemental abundance of Ca for HD 113878 is found to be similar of Ca abundances of the Sun. Other hand, results of JO15 show under-abundances of Ca ion for HD 113878. Both studies are showing under-abundances of Ca ion for HD 118660.

Iron Abundances: - According to the JO15, the elemental abundance of F e for HD 118660 are found to be similar of *Fe* abundances of the Sun. Other hand, results of CA11 show overabundant of F e ion for HD 118660. In the case of HD 113878, under-abundances of Fe ion are reported by CA11 and overabundant of Fe ion is confirmed by JO15.

*Table 6. Elemental Abundances for the HD 113878 and HD 118660 as extracted from the literature.*

| Ion | HD 113878 | | | HD 118660 | | |
|---|---|---|---|---|---|---|
| | JO15 | CA11 | Δ | JO15 | CA11 | Δ |
| Li I | | −8.90 ± 0.13 | | | | |
| C I | −3.97 ± 0.24 | −3.63 ± 0.13 | −0.34 | −4.00 ± 0.00 | −3.50 ± 0.13 | −0.50 |
| Na I | −5.13 ± 0.19 | −5.57 ± 0.14 | +0.44 | −5.50 ± 0.20 | −5.62 ± 0.19 | +0.12 |
| Mg I | −4.69 ± 0.16 | −4.57 ± 0.14 | −0.12 | −4.46 ± 0.01 | −4.59 ± 0.10 | +0.13 |
| Si I | −4.42 ± 0.08 | −4.62 ± 0.09 | +0.20 | −4.49 ± 0.00 | −4.41 ± 0.12 | −0.08 |
| S I | −4.21 ± 0.00 | −4.65 ± 0.10 | +0.44 | | −4.36 ± 0.12 | |
| Ca I | −5.85 ± 0.14 | −5.72 ± 0.13 | −0.13 | −5.78 ± 0.00 | | |
| Ca II | −5.90 ± 0.00 | | | −5.68 ± 0.00 | −5.85 ± 0.10 | −0.17 |

| Ion | HD 113878 | | | HD 118660 | | |
|---|---|---|---|---|---|---|
| | JO15 | CA11 | Δ | JO15 | CA11 | Δ |
| Sc II | −9.57 ± 0.00 | −9.41 ± 0.09 | −0.16 | | −9.00 ± 0.08 | |
| Ti II | −7.20 ± 0.05 | −7.26 ± 0.12 | +0.06 | −7.20 ± 0.05 | −7.37 ± 0.12 | +0.17 |
| V I | −7.84 ± 0.00 | | | | | |
| Cr I | −6.18 ± 0.23 | | | −6.37 ± 0.00 | | |
| Cr II | −5.93 ± 0.20 | −6.26 ± 0.14 | +0.33 | | −6.15 ± 0.03 | |
| Fe I | −4.44 ± 0.10 | | | −4.56 ± 0.04 | | |
| Fe II | −4.39 ± 0.11 | −4.62 ± 0.10 | +0.23 | −4.56 ± 0.04 | −4.10 ± 0.12 | −0.46 |
| Co | −6.29 ± 0.13 | | | | −6.37 ± 0.06 | |
| Ni I | −5.43 ± 0.20 | −5.61 ± 0.12 | +0.18 | −5.73 ± 0.04 | −5.78 ± 0.12 | +0.05 |
| Cu | | −6.00 ± 0.12 | | | | |
| Sr | | −8.33 ± 0.10 | | | −8.89 ± 0.10 | |
| Y II | −8.94 ± 0.29 | −9.10 ± 0.09 | +0.16 | −9.25 ± 0.09 | −9.59 ± 0.06 | +0.34 |
| Zr II | | −8.80 ± 0.06 | | | −8.87 ± 0.08 | |
| Ba II | −8.57 ± 0.00 | −8.56 ± 0.14 | −0.01 | −9.40 ± 0.00 | −9.49 ± 0.07 | +0.09 |
| La II | −9.86 ± 0.05 | | | | | |
| Ce II | −9.16 ± 0.20 | | | | | |
| Nd II | −9.70 ± 0.18 | | | | | |

### 4.3. General Characteristics of Sample Am

The common features of five Am stars are given as below:
1. The heavy rare earth elements (Ni, Y, Zr, Ba, La, Ce, Nd) are showing overabundant property for sample Am stars.
2. The C-peak is shown under-abundant property.

## 5. Stellar Properties

### 5.1. Effective Temperature and Surface Gravity

The effective temperature ($T_{eff}$) of a star is the temperature of a black body of the same size as the star and that would radiate the same total amount of electromagnetic power as emitted by the temperature [50]. According to the comparative result of BTSETTL atmospheric model with spectral energy distribution, the $T_{eff}$ values of HD 73045 and HD 99851 are found to be 7268 K and 6800 K, respectively [51]. Based on principle component analysis (PCA) approach, the value of $T_{eff}$ of HD 73045 comes to be 7199 K [52]. The experimental values of $T_{eff}$ of HD 113878 are 6800 K [51], 7328 K [5], 6900 K [27], 7072 K [26] and 7263 K [53]. The computed values of $T_{eff}$ for HD 118660 are 7772 K [13], 7500 K [53], 7638 K [5], 7177 [51], 7447 [54] and 7340 K [37]. Similarly, surface gravity of a rotating stellar object is the experienced gravitational acceleration at the equator of its surface. In astrophysics, the surface gravity is measured in the logarithm scale of its value in *cgs* system and expressed by log g. The derived and extracted values of $T_{eff}$ and log g through Stromgren photometry, spectroscopic analysis and literature are used to determine the average values of effective temperature and logarithm gravity of each sample Am star. The computed values of $T_{eff}$ and log g of each sample star is given in the Table 8 and these resultant values are used for further analysis of stellar dynamics.

### 5.2. Micro-turbulence Velocity

The microturbulent velocity is defined as the microscale non-thermal component of the gas velocity in the region of the spectral line formation [55]. The theoretical microturbulence velocity ($v_{mic}^{th}$) of each star has determined, using the following relation [56],

$$v_{mic}^{th} = 3.31 \times exp\left[-\left\{\log\frac{T_{eff}}{8071.03}\right\}^2 / 0.01405\right]. \quad (6)$$

The ($v_{mic}^{th}$) values of HD 73045, HD 98851, HD 113878, HD 118660 and HD 207561 are found to be 3.01, 2.63, 2.60, 3.04 and 3.16 km s$^{-1}$ respectively and listed in the Table 8.

***Table 7.*** *The stellar proper motion values through various catalogues.*

| Catalogue | | HD 73045 | HD 98851 | HD 113878 | HD 118660 | HD 207561 |
|---|---|---|---|---|---|---|
| Frouard et al. (2015) [58] | μ$_x$ | −34.26 ± 0.31 | −39.53 ± 0.35 | −22.93 ± 0.27 | 42.92 ± 0.35 | 35.93 ± 0.36 |
| | μ$_y$ | −12.26 ± 0.31 | −16.65 ± 0.35 | −3.96 ± 0.27 | −12.47 ± 0.35 | 28.56 ± 0.36 |
| Perryman et al. (1997, 2009) | μ$_x$ | −35.80 ± 1.19 | −41.91 ± 0.76 | −22.39 ± 0.77 | 41.01 ± 0.79 | 42.92 ± 0.35 |
| | μ$_y$ | −10.58 ± 0.94 | −15.53 ± 0.63 | −5.15 ± 0.65 | −14.98 ± 0.51 | 30.28 ± 0.55 |
| Roeser et al. (2010) [59] | μ$_x$ | −35.70 ± 0.90 | −41.80 ± 0.70 | −22.00 ± 0.70 | 40.60 ± 0.70 | 36.40 ± 0.60 |
| | μ$_y$ | −11.40 ± 0.80 | −15.50 ± 0.60 | −4.90 ± 0.60 | −14.70 ± 0.50 | 30.30 ± 0.60 |
| Van Leewan (2007) | μ$_x$ | −35.71 ± 0.95 | −41.59 ± 0.44 | −22.43 ± 0.67 | −41.01 ± 0.79 | 36.34 ± 0.64 |
| | μ$_y$ | −11.79 ± 0.78 | −16.43 ± 0.37 | −5.40 ± 0.55 | −14.98 ± 0.51 | 30.08 ± 0.45 |
| Zacharias et al. (2013) [60] | μ$_x$ | −35.60 ± 0.50 | −41.60 ± 1.00 | −21.70 ± 0.70 | 40.50 ± 1.00 | 36.40 ± 0.80 |
| | μ$_y$ | −12.50 ± 0.70 | −16.40 ± 1.00 | −5.40 ± 0.80 | −15.50 ± 1.00 | 30.40 ± 0.80 |
| Average Values | μ$_x$ | −35.41 ± 0.77 | −41.29 ± 0.65 | −22.29 ± 0.71 | 41.21 ± 0.73 | 37.60 ± 0.55 |
| | μ$_y$ | −11.71 ± 0.71 | −16.10 ± 0.59 | −4.96 ± 0.57 | −14.53 ± 0.57 | 29.92 ± 0.55 |

### 5.3. Stellar Luminosity

The value of logarithm luminosity of each star has calculated through the relation,

$$\frac{L}{L_\odot} = -\frac{M_V + BC - M_{bol,\odot}}{2.5} \quad (7)$$

The value of $M_{bol,\odot}$ is given as 4.74. After using the values of $M_V$ and BC as estimated in the Section 8.5.1, the values of prescribed logarithm luminosity for HD 73045, HD 98851, HD 113878, HD 118660 and HD 207561 are found to be 1.32, 1.30, 1.53, 1.49 and 1.21 respectively.

### 5.4. Stellar Radius

The stellar parameters $T_{eff}$ of a star is physically related to the total radiant power per unit area at stellar surface (F*) [57]:

$$\sigma T^4 = \int_0^\infty F_\nu d\nu = F_* = \frac{L}{4\pi R^2}, \text{ and } R = R_\odot (\frac{L}{L_\odot})^{1/2} (\frac{T_\odot}{T})^2 \quad (8)$$

It allows to us to estimate the stellar radius by using the values of $T_{eff}$ and luminosity of investigated objects. In this connection, the author used the following relation to estimate the stellar radius,

The value of effective temperature $(T_\odot)$ is 5777 K. The radius of HD 73045, HD 98851, HD 113878, HD 118660 and HD 207561 are computed as 2.77 R⊙, 2.97 R⊙, 3.89 R⊙, 3.34 R⊙ and 2.32 R⊙ respectively. The result of HD 73045 is close to the literature 2.124±0.522 R⊙ [13].

### 5.5. Distance and Stellar Parallax

If, the units of stellar parallax (π) and distance (d) are milli-arcsec (mas) and parsecs (pc), then the stellar parallax is equivalent to the reciprocal of the stellar distance. The listed stellar distances and parallax of Table 1 are extracted from the work of Mcdonald et al [51] and GAIA collaborations respectively. These values are different that of given in the catalogue [29] as given in the Section 3.2. After utilizing these values, author assigned the average parallax values for HD 73045, HD 98851, HD 113878, HD 118660 and HD 207561 as 4.36, 6.24, 2.89, 13.93 and 8.02 mas respectively. These values are listed in the Table 8 with the corresponding values of stellar distances.

### 5.6. Stellar Proper Motions and Transverse Velocities

The proper motion of any star shows the rate of angular drift across the sky. The proper motion values of the sample stars are extracted from the various previous works, as listed in the Table 7. Author computed the proper motion values of each star as an average value of these extracted proper motion values. The proper motion of each star is determined through the following relation,

$$\mu = \sqrt{\mu_x^2 + \mu_y^2} \quad (9)$$

*Table 8. New Computed results for the sample Am Stars.*

| Parameter Name | HD 73045 | HD 98851 | HD 113878 | HD 118660 | HD 207561 |
|---|---|---|---|---|---|
| $T_{eff}(K)$ | 7418 | 7082 | 7059 | 7452 | 7614 |
| $\log g$ | 4.04 | 3.67 | 3.68 | 3.97 | 3.84 |
| $v_{mic}^{th}(km\ s^{-1})$ | 3.01 | 2.63 | 2.60 | 3.04 | 3.16 |
| $\log \frac{L}{L_\odot}$ | 1.32 | 1.30 | 1.53 | 1.49 | 1.21 |
| $E(B-V)$ | -0.012 | -0.37 | -0.413 | 0.183 | 0.75 |
| Mass $(M_\odot)$ | 1.98 | 1.80 | 2.05 | 2.00 | 1.88 |
| Radius $(R_\odot)$ | 2.77 | 2.97 | 3.89 | 3.34 | 2.32 |
| $\mu\ (mas/yr)$ | 37.30 ± 1.05 | 44.32 ± 0.88 | 22.84 ± 0.91 | 43.70 ± 0.93 | 48.05 ± 0.78 |
| $\pi\ (mas)$ | 4.36 | 6.24 | 2.89 | 13.93 | 8.02 |
| $d(pc)$ | 229.36 | 160.26 | 346.02 | 71.78 | 124.69 |
| $v_T(km/s)$ | 8.55 | 7.10 | 7.90 | 3.14 | 5.99 |
| $\tau_{MS}(Gyrs)$ | 14.07 | 13.42 | 14.32 | 14.14 | 13.71 |
| $r_{TL}\ (AU)$ | 0.578 | 0.559 | 0.584 | 0.580 | 0.568 |
| $r_i, r_o\ (AU)$ | 1.095-1.578 | 1.087-1.566 | 1.179-1.677 | 1.164-1.677 | 1.049-1.511 |
| $v_{max}$ | 0.228 | 0.184 | 0.123 | 0.158 | 0.304 |

Author has been found the proper motion values of HD 73045, HD 98851, HD 113878, HD 118660 and HD 207561 as 37.30±1.05, 44.32±0.88, 22.84±0.91, 43.70±0.93 and 48.05±0.78 mas/yr respectively. A perpendicular component of velocity of stellar objects with line of sight, is expressed as the Transverse Velocity $(V_T)$. The value of $(V_T)$ has been obtained as $v_T = 0.21\ \mu D\ km\ s^{-1}$.

In the above relation, units of μ and D are the mas/yr and kpc respectively. This formula leads the values of $V_T$ for HD 73045, HD 98851, HD 113878, HD 118660 and HD 207561 as 8.55, 7.10, 7.90, 3.14 and 5.99 km/s respectively.

### 5.7. Stellar Lifetime

The lifetime of a star on the main sequence can be estimated through the solar evolutionary models. In this connection, the lifetime $(\tau_{MS})$ of a star is expressed as,

$$\tau_{MS} \simeq \left\{\frac{M}{M_\odot}\right\}^{-2.5} \times 10^{10}\ years \quad (10)$$

where M and $M_\odot$ are the stellar and solar masses, respectively. The values of $\tau_{MS}$ of HD 73045, HD 98851, HD 113878, HD 118660 and HD 207561 are found to be 14.07, 13.42, 14.32, 14.14 and 13.71 Gyrs respectively.

### 5.8. Stellar Tidal Locking Radius

An orbital state, where the planet rotates around its own axis with the same speed as it orbits its host star is called 1:1 spin-orbit resonance [61]. To calculate the value of distance up to which a planet would be tidally locked in a time span equal to the age of the Solar system, the following relation of tidal radius is used [62],

$$r_{TL} \approx 0.46 \left(\frac{M}{M_\odot}\right)^{\frac{1}{3}} AU. \quad (11)$$

Applying above relation, the author estimated the value of $r_{TL}$ of HD 73045, HD 98851, HD 113878, HD 118660 and HD 207561 as 0.578, 0.559, 0.584, 0.580 and 0.568 AU respectively.

### 5.9. Stellar Habitable Zones

The circumstellar habitable zone of a star is a range of orbits around it within which a planetary surface can support liquid water given enough atmospheric atmosphere [63, 64]. The approximate inner radii ($r_i$) and outer radii ($r_o$) of the boundaries of the host star's habitable zones are expressed as [65],

$$r_i = \sqrt{\frac{L/L_*}{1.1}}, and\ r_o = \sqrt{\frac{L/L_*}{0.53}} \quad (12)$$

where $r_i$ and $r_o$ are measured in astronomical units (AU). The inner radii of HD 73045, HD 98851, HD 113878, HD 118660 and HD 207561 are estimated as 1.095 AU, 1.087 AU, 1.179 AU, 1.164 AU and 1.049 respectively. Similarly, the outer radii of HD 73045, HD 98851, HD 113878, HD 118660 and HD 207561 are computed as 1.578 AU, 1.566 AU, 1.699 AU, 1.677 AU and 1.511 respectively.

## 6. Correlations Among Stellar Parameters

### 6.1. Stellar Locations in Temperature-Luminosity HR Diagram

The precise location of each star in Hertzprung Russell (HR) diagram effectively useful to know the evolutionary status of a star and positions of present sample stars in temperature-luminosity HR diagram are depicted in the Figure 4. The estimation procedure of most accurate values of effective temperature and luminosity has an error of 150 K in the determination of $T_{eff}$ leads to an error of about 0.20 mag in the bolometric magnitude [26, 66]. The location of the sample stars in the HR diagram consistent with the theoretical work of Turcotte et al. 2000 [67] leads to the young stable Am star against the κ-mechanism pulsation. The positions of HD 98851, HD 113878 and HD 118660 indicates that they become unstable due to their evolution towards the red edge of the instability strip. The location of HD 207561 and HD 73045 in the HR diagram indicates that it is slightly evolved from the main sequence and lies within the δ-Scuti instability strip. Their evolution track follows the path of Henry Track in luminosity-temperature HR diagram. Based on the position of stars in HR diagram, author also estimated the masses of HD 73045, HD 98851, HD 113878, HD 118660 and HD 207561 as 1.99 $M_\odot$, 1.80 $M_\odot$, 2.05 $M_\odot$, 2.00 $M_\odot$ and 1.88 $M_\odot$ respectively.

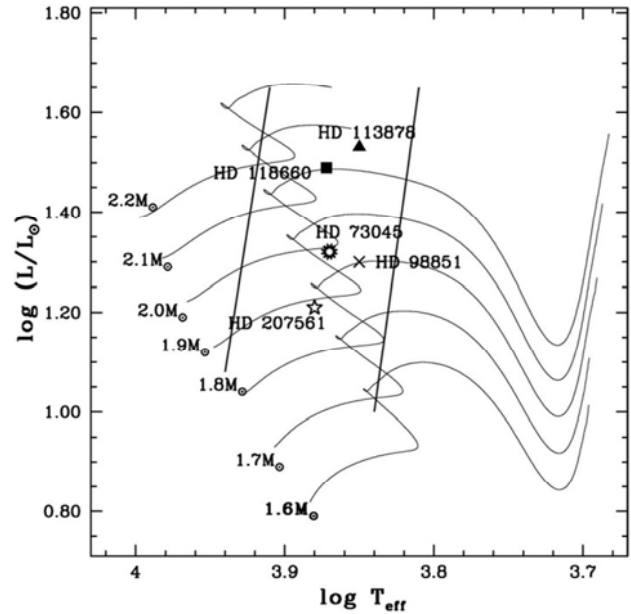

*Figure 4. The location of sample stars in HR diagram.*

### 6.2. An Accuracy Test of Surface Gravity

The surface gravity of all stars can also be derived from the fundamental relation [26]

$$\log\frac{g}{g_\odot} = \log\frac{M}{M_\odot} + 4\log\frac{T_{eff}}{T_\odot} - \log\frac{L}{L_\odot} \quad (13)$$

where L and M are the bolometric luminosity and mass of the studied stars. This relation gives the value of logarithm of surface gravity for HD 73045, HD 98851, HD 113878, HD 118660 and HD 207561 as 3.60, 3.49, 3.47 3.56 and 3.66 respectively. Notice that in above relation mass plays only a secondary role: varying it by 10% changes log g by 0.4 dex [26]. Author has been estimated standard deviation from given values in the Table 8 as 10.9%, 4.9%, 5.7%, 10.3% and 4.7% for HD 73045, HD 98851, HD 113878, HD 118660 and HD 207561 respectively. The values of standard deviation of surface gravity of studied stars confirm the role of mass in the variable value of surface gravity.

### 6.3. Maximum Rotational from the Asteroseismic Mass Scale

Stellar rotation is observed to be a strong function of mass and evolutionary state. The measurement of rotation period could serve as useful diagnostics in the context of asteroseismology [68]. The location of the peak (in power) of envelop of oscillation modes is defined as the frequency of maximum power ($v_{max}$). The value of $v_{max}$ of a star is estimated through the following relation [69],

$$v_{max} = \frac{M/M_\odot}{(R/R_\odot)^2 \sqrt{T_{eff}/T_\odot}} \quad (14)$$

This scaling relation is model independent and an extremely useful to determine the stellar parameters. The values of $v_{max}$ for HD 73045, HD 98851, HD 113878, HD 118660 and HD 207561 are found to be 0.228, 0.184, 0.123, 0.158 and 0.304 d−1 respectively. All stars, having a range of P < 10 days, are either above a solar mass (on the MS or the SGB) or young, rapidly rotating solar mass objects [70]. Thus, the estimated values of the periods of sample Am stars confirm their masses above the solar mass.

## 7. Conclusions

Author obtained a decreasing pattern of stellar magnitudes with an increment of effective wavelengths from filter z to filter K and found a bump of stellar magnitude at filter G. The late A type spectral class of the stars HD 98851 and HD 113878 has confirmed through the results of photometric indices of the sample stars. The mild deviation and nearly solar abundance of elements indicates that they belong to the subclass of marginally Am stars or close to the normal ones as supported by the spectro-polarimetry-measurements of circularly polarized spectra (Joshi et al. 2015). Thus, the abundance analysis of these stars confirms their CP1 (i.e. Am) spectral class. The effective temperature, surface gravity, mass, distance, luminosity, transverse velocity, proper motion and micro-turbulence velocity of selected stars are estimated by author and results are summarized in the Table 8. Furthermore, the surface gravity value (log g) of each star indicates that they are significantly evolved in the HR diagram. The present analysis of the sample stars indicate that the new results of mass and radius are large compared to the previous ones. To see the importance of liquid water to Earth's biosphere, the boundaries ($r_i, r_o$) of habitable zones and tidal radius (($r_{TL}$) has been estimated for each sample star. These values are effectively useful to search the Earth like extra-terrestrial life and intelligence. Since, the calculated stellar lifetime of each studied star is found to be effective low compared to the solar-life time, therefore, their possible revolving planets are not suitable to large human settlement in the terms of greatest time span compare to us. Asteroseismological mass scale test of sample Am stars confirms their mass, having greater than the solar mass.

## Acknowledgements

GCJ is thankful to the Director, ARIES (Nainital) to permit to him to use research work for a thesis entitled "Photometric Analysis of Open Star Clusters and Variability of Stars" through the letter No. AO/2018/41 on date 12 April 2018. GCJ is also thankful for Vice-Chancellor, Kumaun University to permit to him to include the authorized research work in the revised thesis draft through Letter No. KU/VC/SRICC/2018/50/ date 24-01-2018 and Letter No. SRICC/2018/Gireesh-Shodh/4500 date 23 October 2018. GCJ is also acknowledged SIMBAD, VIZIER services and ESO. GCJ is also giving especial thanks to Dr. Eugene Semenko (National Astronomical Research Institute of Thailand) for transferring Figure 2 to the corresponding author via Dr. Santosh Joshi during construction of Joshi et al. (2015).

International Journal of Astronomy, Astrophysics and Space Science 2019; 6(3): 25-37                                    37